\documentstyle[sprocl]{article}

\def\NPB#1#2#3{Nucl. Phys. B {\bf#1} (19#2) #3}
\def\PLB#1#2#3{Phys. Lett. B {\bf#1} (19#2) #3}
\def\PRD#1#2#3{Phys. Rev. D {\bf#1} (19#2) #3}
\def\PRL#1#2#3{Phys. Rev. Lett. {\bf#1} (19#2) #3}
\def\PRT#1#2#3{Phys. Rep. {\bf#1} (19#2) #3}
\def\DVN{D.~V.~Nanopoulos}

\def\be{\begin{equation}}
\def\ee{\end{equation}}
\def\bea{\begin{eqnarray}}
\def\eea{\end{eqnarray}}

\begin{document}
\begin{flushright}
ACT-12/96\\
CTP-TAMU-43/96\\
{\tt hep-ph/9608489}
\end{flushright}

\title{Theoretical Summary\footnote{XI Topical Workshop on Proton-antiproton
Collider Physics, Abano Terme (Padua), Italy, 26 May--1 June, 1996.}}

\author{D. V. Nanopoulos}

\address{Center for Theoretical Physics, Department of Physics, Texas A\&M
University, College Station, TX 77843-4242, USA}
\address{Astroparticle Physics Group, Houston Advanced Research Center,
The Mitchell Campus, 4800 Research Forest Drive, The Woodlands, TX 77381, USA}

\maketitle\abstracts{
After a quick tour through the present status of the Standard Model, an
attempt is made to set up a framework to discuss some presently available
exotica, including $R_b$ and the CDF $ee\gamma\gamma+E_T\hskip-13pt/\ $
``event." Supersymmetry seems to be a key player in establishing a paradigm
shift beyond the Standard Model.}

\section{Introduction}
My task here is twofold: to summarize the theory talks given at this very
exciting workshop {\em and} to provide some discussion on some theoretical
aspects of supersymmetry, supergravity, and superstrings in a simplified way.
Clearly, this seems virtually impossible, if it were not for the high quality
and clarity of the talks we heard, and by the fact that supersymmetry has been
one of the main experimental issues in this workshop. In the next section I
will discuss some topics within the Standard Model, while in Section 3 I will
provide some possible evidence/hints entailing an extension of the Standard
Model. Section 4 is devoted to some characteristics of the Superworld, paving
the way to some possible interpretations of presently existing exotica, such
as $R_b$ and the CDF $ee\gamma\gamma+E_T\hskip-13pt/\ $ ``event", discussed
in Section 5.

\section{Standard Model Forever (?)}
As we have repeatedly heard in this workshop, the Standard Model (SM) seems to
be alive and in very good shape. More and more sectors of the SM get probed
experimentally and still the results start to sound monotonous: no problem!
{}From LEP1/1.5, to FNAL collider to HERA to ... all the data seem to be
unanimously in favor of the SM. Sometimes one wonders what is so special about
an $SU(3)_C\times SU(2)_L\times U(1)_Y$ gauge theory with three generations
of quarks and leptons and spontaneous electroweak breaking (by the Higgs
mechanism?) to describe nature so well. And to think that no theorist worth
her soul wants it to be valid because of its numerous well-known defaults: too
many parameters, no ``real" unification, no gravity in sight, etc, etc. As
Hollik~\cite{1} told us, the Electroweak (EW) precision observables, not
only fit within the SM, but their sensitivity, through radiative corrections,
to the mass of the Higgs boson, enables us to put an upper bound of $M_H\le
300$~GeV at 95\%CL. After the successful prediction of the top-quark mass
\cite{2}, EW precision physics provides once more valuable information.
While clearly EW precision data is the defining factor of the experimental
worthiness of the Standard Model, all other information is in excellent
agreement with the SM. As discussed here by Pilaftsis~\cite{3}, the
weak sector of the SM, related to the Cabibbo-Kobayashi-Maskawa (CKM) weak
mixing matrix, including CP violation, seem to be in good phenomenological
shape, amid the fact that we may not have as yet a universally acceptable
theoretical explanation of the fermion mass spectrum, their mixings as well as
of the origin of CP violation. Many of us believe that the ``flavor problem"
will find its solution at high energies/short distances, involving ``new
physics" beyond the Standard Model.

The status of CPT symmetry was also discussed here~\cite{3} rejecting,
prematurely in my opinion, any possible framework that may lead to CPT
violation. Since this is (or could be) a matter of the upmost importance,
let me digress a bit. In point-like Quantum Field Theory one can prove, based
on general principles like locality, Lorentz invariance, and unitarity,
that CPT is always conserved, the well-known CPT theorem~\cite{4}. On the other
hand, in Quantum Gravity one may see one or more of the above general
principles not to hold, thus giving a chance to CPT violation. Specifically, in
string theory, the only known framework for a consistent theory of Quantum
Gravity, this possibility has been raised~\cite{5}, and experimental efforts,
mainly by the CPLEAR Collaboration~\cite{6}, are reaching a very exciting
range where some form of CPT violation may be observable. Namely, while the
CPLEAR Collaboration determines that~\cite{6} $m_{K^0}-m_{\bar
K^0}/m_{K^0}\le9\times 10^{-19}$, naive quantum gravitation expectations lead one to believe that
\begin{equation}
{m_{K^0}-m_{\bar K^0}\over m_{K^0}}\sim {m_{K^0}\over
M_{Pl}}\sim(10^{-18}-10^{-19})
\end{equation}
with $M_{Pl}=(\sqrt{G_N})^{-1}$, with $G_N$ the gravitational constant. It may
be that these naive expectations, elaborated further and supported by some
string calculations~\cite{5}, are too optimistic and that the RHS of this
equation is further suppressed, but stilll it is a very worthwhile effort
and {\em we have to know} the best possible limits to CPT violation. After all,
introducing an {\em arrow of time} in microscopic physics, is not a marginal
issue.

The status of Quantum Chromodynamics (QCD) is not in bad shape either. For
example, as discussed here by Mangano~\cite{7}, rather elaborate perturbative
QCD calculations, including resummation effects, take care of the bulk of the
experimental data on jet physics, presented abundantly at this workshop. To my
relief, the so called ``CDF anomaly", referring to the CDF observation~\cite{8}
that the high statistics inclusive jet production measurements at the Tevatron
imply a larger cross section at high jet $E_t$ than expected from NLO QCD
calculations, seem to be accountable within the standard QCD framework
\cite{7,9,10}. Indeed, both the CDF and D0 inclusive jet cross sections are
found to be in good agreement using a uniform theoretical NLO QCD calculation,
taking into account the different kinematic coverages of the pseudo-rapidity
variable ($\eta$) in the two experiments~\cite{9}. Subtle effects of
jet algorithms, scale-choice and delicate cancellations amoung various
contributions needed to be handled with extra care in the NLO QCD calculations,
if precision down to the few percent level is required. Anyway, this is a very
delicate analysis, still in progress, and we may have not yet heard the final
verdict, but it looks like there is no need to call for ``new physics" in order
to explain the observed excess.

Further evidence for the validity of perturbative QCD calculations, including
resummation effects, was presented by Berger~\cite{11}, concerning this time
the inclusive cross section for the production of top quark-antiquark ($t\bar
t$) pairs in hadron reactions. A measure of the success of these calculations
is the remarkable agreement between the theoretical predictions, worked out
independently by three groups: (for $m_t=175$~GeV and $\sqrt{s}=1.8$~TeV)
\begin{equation}
\sigma^{\rm th}_{t\bar t}[{\rm pb}]=\left\{\begin{array}{ll}
4.95^{+0.70}_{-0.40}&[12]\\
5.52^{+0.07}_{-0.42}&[13]\\
4.75^{+0.63}_{-0.68}&[14]\end{array}\right.
\end{equation}
and the Tevatron measurements:
\begin{equation}
\sigma^{\rm exp}_{t\bar t}[{\rm pb}]=\left\{\begin{array}{lll}
7.6^{+1.9}_{-1.5} &{\rm(CDF)}&[15]\\
5.2\pm1.8 &{\rm(D0)}&[16]\end{array}\right.
\end{equation}
On the non-perturbative QCD front, as we heard from Del Duca~\cite{17}, quite
remarkable progress has been made in trying to comprehend the vast amount of
new data on diffraction, presented at this workshop. It looks to me that this
is a very promising field of research, picking up momentum and maturing
rapidly. Several other measurements at the Tevatron on different processes
including W/Z production cross sections, W/Z+jets, Drell-Yan, W charge
asymmetry, Z forward-backward asymmetry, double boson production, photon
production, seem to fit nicely within the Standard Model.

All in all, the agreement between the present experimental status and the SM is
rather remarkable and makes one wonder why we have to go ...

\section{Beyond the Standard Model}
There are many theoretical reasons indicating that an extension of the Standard
Model is unavoidable. Non-inclusion of gravity in the unification program makes
it incomplete. Here though, I would like to take a more pragmatic,
close-to-the-experiment approach and argue that we have already experimentally
strong hints that we have to enlarge the Standard Model. I am not referring
here to the possible ``exi(s)ting exotics" that may provide the first real
cracks of the SM, as will be discussed below, but instead put together a
convincing (?) case from within the Standard Model. What I am referring to here
is to two {\em tentative} indications from precision (mainly LEP) data. The
first one concerns the Higgs boson, that as we have heard here~\cite{1} (see
also~\cite{2}) is probably ``light" ($\le300$~GeV). Actually, a recent global
fit~\cite{2}, including all available data gives
\begin{equation}
M_H=145^{+164}_{-77}\,{\rm GeV}\quad{\rm and}\quad m_t=172\pm6\,{\rm GeV}
\end{equation}
a rather ``light" Higgs boson indeed! While these are wonderful news for
supersymmetric models, which generally predict
\begin{equation}
m^{\rm susy}_h\approx M_Z\pm40\,{\rm GeV}\ ,
\end{equation}
it disfavors strongly-interacting Higgs scenarios, such as technicolor and
the likes.

The second clue refers to the measured values of the three gauge coupling
constants, indicating unification at a very high energy scale ($M_{\rm
LEP}\approx10^{16}\,{\rm GeV}$) and strongly favoring~\cite{18} supersymmetric
GUTs, while excluding non-supersymmetric GUTs, which were already in deep
trouble by the strict proton decay stability limits~\cite{19}. As a
quantitative measure of the above statements we may use the predictions of
GUTs and supersymmetric GUTs concerning $\sin^2\theta_W$ and confront them
with the corresponding experimental value
\begin{equation}
\sin^2\theta_W |^{\rm exp}_{M_Z}\approx 0.232;\quad
\sin^2\theta_W |^{\rm no susy}_{M_Z}\approx0.203;\quad
\sin^2\theta_W |^{\rm susy}_{M_Z}\approx0.230
\end{equation}
The numbers speak for themselves! One may want to add the successful prediction
of the $m_b/m_\tau$ ratio and its straightforward implication of $N_\nu=3$,
\cite{20} vindicated by LEP measurements, as further support for supersymmetric
unification. Furthermore, the amazing success of the Standard Model severely
constrains any attempt to extend it since new dynamical degrees of freedom may
mess up its rather delicate structure. In this case too, supersymmetric models
manage to escape unscathed, in sharp contrast with dynamical symmetry breaking
models, that as we have heard in this workshop, by De Curtis and Chiappetta
\cite{21} may have to just walk or limp or ...

So if we look at the score board, supersymmetric models do pretty well
across-the-board, while dynamical symmetry breaking models leave too much
to be desired. Thus, it is no accident or shouldn't be that surprising that
the supersymmetric extension of the Standard Model has become the major
framework for analyzing ``new physics". In the late seventies, two major
schools of thought were formed, one believed that ``elementarity" of quarks,
leptons, gauge bosons, etc continues all the way (or) close to the Planck
length ($\ell_{Pl}\sim10^{-33}$~cm), while the other school believed that
``hell breaks loose" at the Fermi length ($\ell_F\sim10^{-16}$~cm) and
``elementarity" of at least some of the Standard Model particles has to be
given up. I strongly believe that we have gathered enough evidence supporting
the first point of view, {\em, i.e.}, the fundamental constituents of the
Standard Model keep their ``elementarity" up to very short distances, thanks
to ...

\section{The Superworld}
Since the rudiments of supersymmetry (SUSY) have been discussed rather
extensively in this workshop by our experimental colleagues while presenting
discovery limits on different SUSY particles, I will concentrate here on some
issues bearing direct consequences on the ``exi(s)ting exotics" to be discussed
in the next section. From a pragmatic/practical point of view, the role of
supersymmetry may be effectively described as a {\em gauge hierarchy
stabilizer}, thus leading to the following well-known relation
\begin{equation}
\widetilde m^2\equiv m^2_B-m^2_F\approx {\cal O}(0.1-1\,{\rm TeV})^2
\label{eq:6}
\end{equation}
where $\widetilde m$ represents the mass splitting between the fermion and
boson in the same supermultiplet, i.e., it is a characteristic, generic,
SUSY breaking scale, while the RHS represents the quantitative statement of
the gauge hierarchy stability. It is because of this rather fundamental role
of SUSY as a hierarchy stabilizer (\ref{eq:6}) that the hope exists for a
whole new world, that of the superpartners of `our world', to be within the
discovery potential of presently available or soon to exist accelerators. A
new issue then arises, that of the SUSY breaking mechanism, i.e., who provides
the seeds for SUSY breaking and why is  the SUSY breaking scale ($\widetilde
m$) ${\cal O}(M_W)$? Before we move further, it is worth recalling that because
the defining anticommutator of the SUSY generators ($\{Q,Q\}\propto P_\mu$)
provides the four-momentum generator ($P_\mu$), and since the latter is
involved in General Relativity, we automatically get {\em local SUSY} or
{\em Supergravity}. The only way to break a local symmetry, consistently, is
spontaneously, thus we are led to consider spontaneous SUSY breaking. The form
and structure of supergravity interactions is such that spontaneous SUSY
breaking is achievable in some sector of the theory, let us call it the {\em
Hidden Sector} (H), and it is transmitted through the ubiquitous gravitational
interactions, playing here the role of the {\em Messenger Sector} (M), to the
{\em Observable Sector} (O) of the known quarks, leptons, gauge bosons, Higgs
bosons, etc.~\cite{22}. Minimalistic applications of the above scenario
routinely give
\begin{equation}
m_{3/2}\approx m_0\approx m_{1/2}\approx {\cal O}(\widetilde m)\approx
{\cal O}(0.1-1\,{\rm TeV})
\label{eq:7}
\end{equation}
where $m_{3/2}$ is the mass of the {\rm gravitino}, the spin-3/2 superpartner
of the spin-2 graviton, while $m_0$ and $m_{1/2}$ are the {\em primordial}
seeds of SUSY breaking in chiral multiplets (e.g., $q-\tilde q$,
$\ell-\tilde\ell$, $h-\tilde h$, ...) and in gauge multiplets (e.g.,
$\gamma-\tilde\gamma$, $W-\widetilde W$, $Z-\widetilde Z$, $g-\tilde g$, ...)
respectively. Usually $m_0$ and $m_{1/2}$ get renormalized through strong/electroweak interactions before they yield the experimentally measurable SUSY spectrum, generically represented here by $\widetilde m$. The gravitino, the gauge fermion of local SUSY, becomes massive by absorbing a spin-1/2 fermion, the {\em Goldstino}, through the super-Higgs mechanism, analogous to the usual Higgs mechanism of gauge theories. Phenomenologically, without asking too many questions at the microscopic level, such a generic picture as presented above, has met with considerable success in the following sense. It survived all the severe experimental tests, it succeeded to {\em reproduce} the Standard Model, without at the same time pushing SUSY masses to very high values, thus still experimentally observable, while {\em naturally} leading to gauge coupling unification `observed' at LEP~\cite{18}. A characteristic experimental signature of SUSY has been missing $E_T$. Indeed, in the standard SUSY framework SUSY particles are produced always in pairs, and thus there is {\em always} a lightest supersymmetric particles (LSP), that is stable and escapes the detector. Usually the LSP is identified with the {\em neutralino}
($\chi^0_1$), a linear combination of the electroweak neutral gauginos and the
higgsinos~\cite{23}. The LSP is considered today as one of the main candidates
for the Dark Matter (DM) of the Universe, as explained to us here by Turner
\cite{24}. Thus the experimental signature of SUSY, missing $E_T$, may be
summarized as a {\em dark signal}.

	While this minimalistic point of view has been quite successful in
deriving phenomenologically viable SUSY models, it is characterized by several
drawbacks. To start with, a gravitino mass in the mass range indicated in
(\ref{eq:7}) and dictated by the resolution of the gauge hierarchy problem
is just in the middle of the {\em cosmologically forbidden region}, causing
unacceptable modifications to the primodial nucleosynthesis program
\cite{25,24}. Furthermore, while in the SUSY framework presented above it is
possible to understand {\em dynamically}, the electroweak breaking caused by SM
radiative corrections and ``derive" that $M_W\sim e^{-1/``\alpha"}M_{Pl}$,
where ``$\alpha$" is sone calculable function of gauge and top-quark Yukawa
couplings~\cite{22}, the correlation $\widetilde m\approx{\cal O}(M_W)$
remains a mystery, and looks like another hierarchy problem! In addition, one
has to resort to {\em extraneous} fine-tuning to banish the cosmological
constant ($\Lambda_c$) {\em even at the classical (tree) level}, while one has
to fight hard the menace of SUSY FCNC~\cite{26}, and of dimension-5 operators,
endemic in SUSY theories, causing very fast proton decay \cite{27}, not unrelated with the
fine-tuned way that the Higgs pentaplet gets split into a colored triplet and
a weak doublet. While the above objections may look to the eyes of some
experimentalists or hard phenomenologists as superfluous fine-printing, it
looks to me like very important guiding principles, that may be able to
navigate us to the right model. Let me remind you that the absence of FCNC in
gauge theories, as exemplified by the tiny rate for $K_L\to\mu^+\mu^-$, while
it looked of marginal importance to many, it did not so for Glashow,
Iliopoulos, and Maiani, who tried to understand it naturally, thus introducing
charm, and the rest is history.

There is a specific type of supergravity {\em no-scale supergravity}~\cite{22},
that may hold the key to the solution of many of the above mentioned
conundrums. It has been discovered by its defining property of providing {\em
naturally}, without any fine-tuning, a vanishing cosmological constant
$\Lambda_c$, at the classical level, {\em after} spontaneous SUSY breaking
\cite{28}. Furthermore, it has been used to {\em dynamically} determine,
through SM radiative corrections, that $\widetilde m\approx{\cal O}(M_W)$,
thus dynamically justifying (\ref{eq:6}) and thus completing the SUSY solution
to the gauge hierarchy problem~\cite{29}. In addition, a large class of
no-scale supergravity models~\cite{30}, possessing a global, non-compact
SU(N,1) symmetry, endemic in extended supergravities, is characterized by an
{\em effective decoupling} of the local SUSY breaking scale from the global
SUSY breaking scale. In other words, it is possible to have local SUSY
breaking, while at the same time global SUSY is unbroken, thus in principle
enabling us to drop the $m_{3/2}$ term from (\ref{eq:7}). More specifically,
in this class of models one gets {\em dynamically}~\cite{30}
\begin{equation}
m_0=A_0=B_0=0
\label{eq:8}
\end{equation}
where $A_0$ and $B_0$ refer to the Yukawa and Higgs SUSY breaking interaction
terms. Thus, according to (\ref{eq:7}), the whole burden for global SUSY
breaking is placed on $m_{1/2}$, and indeed very interesting models have
been constructed realizing this picture. Actually, the {\em dynamically}
derived universality (\ref{eq:8}) leads to an automatic resolution of the SUSY
FCNC problem, since the squark and slepton masses are generated mainly through
SM gauge couplings of the superpartners and thus are the same for, say $\tilde
u$ and $\tilde c$, all proportional to the universal $m_{1/2}$! In other words,
the down squark mass matrix is proportional to the unit matrix and thus
diagonal in any basis, including that one that diagonalizes the down quark mass
matrix, thus enabling us to pass on the natural absence of FCNC in gauge
theories to SUSY theories.

The effective decoupling between local and global SUSY breaking scales, as
emerges naturally in no-scale supergravity, has led to a very entertaining
possibility, namely that of a Very Light Gravitino (VLG)~\cite{31}. Indeed,
in a certain class of no-scale models one can show that the following relation
holds~\cite{31}
\begin{equation}
m_{3/2}\approx \left({m_{1/2}\over M}\right)^p M
\label{eq:9}
\end{equation}
where $M\approx10^{18}\,{\rm GeV}$ is the appropriate gravitational scale.
For ${3\over2}\le p\le2$ and $m_{1/2}\approx{\cal O}(100\,{\rm GeV})$ as it
is dynamically determined, one gets
\begin{equation}
10^{-5}\,{\rm eV}\le m_{3/2}\le 1\,{\rm KeV}
\label{eq:10}
\end{equation}
a rather light gravitino indeed. Interestingly enough, the mass range
(\ref{eq:10}) lies {\em outsise} the cosmologically forbidden region~\cite{25},
thus there is no embarrassment in dealing with primordial nucleosynthesis.
Another puzzle gets resolved. Nevertheless, such a very light gravitino has
far-reaching experimental consequences, as first emphasized by Fayet~\cite{32}.
In a nutshell, in interactions involving the gravitino, or more correctly its
longitudinal spin-1/2 component the Goldstino, one has to replace the
gravitational constant $G_N$ by $G_N(\widetilde m/m_{3/2})^2$, thus
effectively transmutting gravitational interactions into weak interactions for
a large fraction of the mass range (\ref{eq:10})! Such a gigantic enhancement
of the gravitino interactions is bound to have a lot of experimental
consequences. Interestingly enough, most of the mass range (\ref{eq:10}) is
still phenomenologically admissible. The main characteristic of a VLG scenario
is that undoubtedly, in this case, the gravitino is the LSP. As emphasized
in~\cite{31}, the neutralinos ($\chi^0_1$) are unstable, decaying mainly to
photons ($\gamma$) and the gravitino, with a lifetime proportional to
$(m^5_{\chi^0_1}/(M m_{3/2})^2)^{-1}$, thus depending on the gravitino mass,
and offering the possibility of neutralino decay inside the detector. In such a
case, the new experimental signature of VLG SUSY is $\gamma$'s plus missing
energy ($E_T\hskip-13pt/\quad$), in other words a {\em light signal}, in sharp
constrast with minimalistic SUSY where we expect (as discussed above) a
{\em dark signal}! In the case of VLG SUSY one has to resort to other particles
(instead of the neutralino) to provide the dark matter in the universe, as
discussed in this workshop~\cite{24}.

It should be emphasized that the no-scale supergravity framework can accomodate
any type of gravitino, from superheavy ($m_{3/2}\sim{\cal O}(M)$)~\cite{33},
to minimalistic ($m_{3/2}\sim{\cal O}(M_W)$)~\cite{29,30}, to very light
($m_{3/2}\sim (m_{1/2}/M)^p M$)~\cite{31}, in each case providing a rather
constrained, highly economical (in terms of free parameters), and
experimentally falsifiable model. While no-scale supergravity seems to resolve
several of the drawbacks of the minimalistic viewpoint discussed above, due to
its specific structure, the question has been frequently raised about the
stability of the no-scale structure, when quantum corrections are taken into
account. While {\em naively} the no-scale structure seems to collapse, we
have retained for years the hope that such an amazing and rich structure should
may perhaps be an exact property in the ``right" quantum theory of gravity.
Our hopes were not an illusion! Indeed, string theory, the only known
consistent quantum theory of gravity, seems to yield as its long wavelength
limit SU(N,1) no-scale supergravity, as first proven by Witten~\cite{34}. This
``derivation" was valid in the weak coupling limit of string theory, thus once
more the quantum stability of no-scale supergravity was in doubt. Lo and
behold, during the last few months things have changed dramatically. String
dualities, believed to be {\em exact} symmetries, have provided us with very
powerful tools to map strongly-coupled string theories to weakly coupled ones.
Specifically, the $E_8\times E'_8$ heterotic string theory, which in its weak
coupling yields~\cite{34} no-scale supergravity, has a strong coupling limit
dual to the 11-dimensional long-wavelength limit of ``M-theory", which has
been very recently proven, by Banks and Dine~\cite{35} and Horava~\cite{36},
to yield, within some controllable approximations, nothing else by no-scale
supergravity! In other words, $E_8\times E'_8$ heterotic string theory keeps,
basically intact, the no-scale structure all the way from the weak coupling to
the strong coupling limit, i.e., {\em including all quantum corrections}, to
some controllable approximation. It does not take an heroic effort to dare to
suggest that like string duality, the no-scale structure is an {\em exact}
string property, far beyond the limits of perturbation theory that has been
discovered, or ``derived" in string theory, and thus eventually leading to a
clear understanding of the natural vanishing of the cosmological constant
($\Lambda_c$) {\em exactly}, and not merely at the classical level.

Beyond high-brow theoretical consequences, these recent developments involving
``M-theory" may have rather drastic and far-reaching
phenomenological/experimental consequences, giving an unforseeable twist to the
whole unification and SUSY model building program. Here is a micrography of
what is going on. The 11th dimension, which becomes the 5th dimension after
suitable compactification, seems to play a very peculiar and unheard before
role. The extra 5th dimension, instead of being as usual periodic, it is a
segment,~\cite{37} with the gauge and matter fields living at the endpoints
only, while the supergravity and moduli fields propagate in the
five-dimensional bulk! That is, spacetime is a {\rm narrow} five-dimensional
layer bounded by four-dimensional walls.  At the one end ``live" the
``{\em observable fields}" (quarks, leptons, gauge bosons, etc.) coming from
one of the $E_8$'s, while at the other end ``live" the ``{\em hidden or shadow
fields}" contained in the other $E_8$. It is remarkable that when the coupling
constants of the observable sector get their ``normal values" (e.g.,
$\alpha_{\rm GUT}\sim1/25$) the hidden sector ($E'_8$) coupling constant is
driven to its strong limit, enabling it to form a gaugino condensate, a
prerequisite for spontaneous local SUSY breaking. Actually, for distances
$\ell$ between the ``normal" 6-dimensional compactification radius $R_{KK}$
(i.e., 10D$\to$4D) and the 5-dimensional compact radius $R_5$ (i.e.,
$R_{KK}\le\ell\le R_5$) even if the gaugino condensate has been formed, there
is no local SUSY breaking. For distances $\ell>R_5$, local SUSY breaking occurs
and clearly $m_{3/2}=f(R_5,...)$, such that $m_{3/2}\to0$ as $R_5\to\infty$.
Clearly, the 5th-dimension protects local SUSY, and the ``geometrical picture"
above is very suggestive and explains nicely the natural emergence of the
no-scale structure in ``M-theory".

It should not escape our attention the fact that the {\em scheme} discussed
above: hidden$\to$messenger$\to$observable sector, for the transmission of
SUSY breaking, is {\em literally} reproduced here. The one four-dimensional wall containing the $E'_8$ is the {\em hidden sector}, the five-dimensional bulk with the supergravity, moduli fields is the {\em messenger sector}, and the other four-dimensional wall contains the {\em observable sector}. What is
surprising is the fact that the onset of the fifth dimension leaves the
observable sector intact. Gauge, Yukawa, and scalar interactions of the
Standard Model are {\em oblivious} to the existence of the fifth dimension!
This observation was made by Witten~\cite{38}, who suggested that if the fifth
dimension is suitably turned on below the unification scale ($M_{\rm
LEP}\sim10^{16}\,{\rm GeV}$), it may provide a kink in the gravitational
coupling so that all couplings meet at $M_{\rm LEP}\sim10^{16}\,{\rm GeV}$
(i.e., by $G_N E^2\to G_N E^3...$), thus resolving the possible problem arising
because of the disparity between $M_{\rm LEP}$ and the weak coupling string
limit $M_{\rm string}\sim5\times10^{17}\,{\rm GeV}$. Geometrical/topological
decoupling between observable/supergravity/hidden fields in ``M-theory" is
suspiciously reminiscent of the decoupling that occurs naturally in SU(N,1)
no-scale supergravity between local and global SUSY breaking, discussed above,
and thus even if the ``M-theory" is still in its infancy, it is not inconceivable that some formula similar to (\ref{eq:8}),(\ref{eq:9}) may eventually pop up from ``M-theory". Further indirect evidence in support of such a viewpoint has been provided in Ref.~\cite{36}, where it has been shown that the role of the would-be goldstino, to be absorbed by the gravitino in its way to becoming massive and thus breaking local SUSY, is played by the {\em normal component} of the 11-dimensional gravitino! In a way ``M-theory" provides an effectively sealed, from the ``observable sector", local SUSY breaking mechanism where all the ingredients are ingeniously provided by the 11th (eventually becoming the 5th) dimension.

It is amusing to notice that very early (pre string-theory) attempts~\cite{39}
to make sense out of D=11,N=1 supergravity theories suggested a gravitino
mass of the form $m_{3/2}\sim(m_{1/2}/M)^2 M$, i.e., of the form given by
(\ref{eq:9}) with $p=2$! Actually, the 5-dimensional gravitational constant
$G_N^{\rm 5-D}$ seems to be much larger that the ``normal" 4-dimensional one
$G_N$, making one wonder whether the effective replacement in the case of VLG,
of $G_N$ by $G_N(\widetilde m/m_{3/2})^2$ discussed above, is somehow related
(i.e., $G_N^{\rm 5-D}=f[G_N(\widetilde m/m_{3/2})^2]$. To put it {\em bluntly}:
is the VLG scenario the macroscopic ``tip" of a microscopic ``M-theory" 5th
dimension? We don't know yet, but we are very likely going to know soon.

While we have tried to provide solutions to most of the drawbacks of the
minimalistic SUSY framework, the puzzle of the dimension-five proton decay
operator has not been dismissed. Actually, this problem gets worse if something
like (\ref{eq:8}) is valid, because we do not have enough free parameters to
play around, and specifically $m_0/m_{1/2}>{\cal O}(3)$ is required~\cite{40}.
String theory comes once more to our rescue. For many well-known reasons
\cite{41}, an SU(5)$\times$U(1) unified gauge theory is most favored in string
theory. The similarity to SU(2)$\times$U(1) should be obvious, as the less
known fact that this the {\em only known} string theory where fractional
electric charges (e.g., $\pm1/2$, etc.) get automatically confined in a way
resembling $\rm SU(3)_{\rm color}$ (QCD).~\cite{42}

The defining property of SU(5)$\times$U(1) is that it reshuffles quarks and
leptons in a {\bf10} and $\bar{\bf5}$ in a way
\begin{equation}
{\bf10}=\left\{\left(\begin{array}{c}u\\
d\end{array}\right),d^c,\nu^c\right\}\quad;
\bar{\bf5}=\left\{u^c,\left(\begin{array}{c}\nu_e\\
e\end{array}\right)\right\}\quad;{\bf1}=e^c
\label{eq:11}
\end{equation}
different from SU(5)
\begin{equation}
{\bf10}=\left\{\left(\begin{array}{c}u\\
d\end{array}\right),u^c,e^c\right\}\quad;
\bar{\bf5}=\left\{d^c,\left(\begin{array}{c}\nu_e\\ e\end{array}\right)\right\}
\label{eq:12}
\end{equation}
by {\em flipping} $u^c\leftrightarrow d^c$ and $e^c\leftrightarrow\nu^c$ and
thus the reason some call SU(5)$\times$U(1) {\em flipped} SU(5). By making the
{\bf10} contain an SU(3)$\times$SU(2)$\times$U(1) singlet ($\nu^c$), it makes
it useful (in the Higgs version) to break SU(5)$\times$U(1) directly down to
SU(3)$\times$SU(2)$\times$U(1), without the use of adjoint representations,
thus getting the blessing of string theory. The structure (\ref{eq:11}) leads
also to a natural Higgs-triplet-doublet splitting, resolving thus another
minimalistic SUSY puzzle, while at the same time banishing the dangerous
dimension-five proton-decay operators! Thus, we were led to consider~\cite{43}
a stringy, no-scale, SU(5)$\times$U(1) theory obeying (\ref{eq:8}), but still
free of d=5 proton decay operators, either in its minimalistic form~\cite{43}
(i.e., satisfying (\ref{eq:7})) or in its VLG form~\cite{44} (i.e., satisfying
(\ref{eq:9})). In order to find out which way nature prefers if any in the
rather broad framework developed in this section, we have to pay some due
attention to the presently available ...

\section{Exi(s)ting Exotics}
Most of the experimental and theoretical talks in this workshop finished with
the, by now, expected words to the effect ``we have seen nothing unusual or
unexplainable by the Standard Model ...". Fortunately there are two very
noticeable exceptions, that of $R_b$ and the CDF
$ee\gamma\gamma+E_T\hskip-13pt/\ $ ``event". Let me discuss each of these
in turn.
\subsection{$R_b$}
It is by now well-known that if we define $R_Q\equiv\Gamma(Z^0\to Q\bar
Q)/\Gamma(Z^0\to{\rm hadrons})$, then we have the following theoretical(SM)-
experimental mismatch for $R_b$ and $R_c$:
\begin{equation}
R^{\rm exp}_b=\left\{\begin{array}{ll}
0.2202\pm0.0016,&{\rm for}\ R^{\rm SM}_c=0.172\\
0.2211\pm0.0016,&{\rm for}\ R_c\ {\rm``free"}\end{array}\right.
\label{eq:13}
\end{equation}
while $R^{\rm SM}_b=0.2157$, and
\begin{equation}
R^{\rm exp}_c=0.160\pm0.007
\label{eq:14}
\end{equation}
while $R^{\rm SM}_c=0.172$. 

What is even more peculiar is the fact that the
leptonic Z-widths ($\Gamma_{\rm lep}$) and the {\em total} hadronic Z-width
($\Gamma_{\rm had}$) seem to be in very good agreement with the SM ($|\Delta
\Gamma_{\rm had}|<3\,{\rm MeV}$). One of course may take the attitude that
$R^{\em exp}_{b,c}$ are just some experimental flukes/fluctuations and they
will eventually ``relax" to their SM values. Something perhaps already
happening, with at least $R^{\rm exp}_c$! On the other hand, as discussed in
considerable detail by Feruglio~\cite{10}, we can use the so-called
``$R_{b,c}$-crisis" to see how well we are doing with the extensions of the SM
discussed in the previous sections. The first thing that comes to mind is of
course SUSY contributions. Indeed, it has been suggested~\cite{45} that SUSY
loop corrections to the $Z-b-\bar b$ vertex involving ``light" charginos
($\chi^\pm_1$) and top squarks ($\tilde t$) may provide a $\Delta R^{\rm
susy}_b$ such as to close the gap between theory and experiment, as indicated
in (\ref{eq:13}). While at first sight this statement sounds plausible, things
get a bit more complicated. If we take into account {\em all} available
constraints from: LEP~1.5 limits on chargino ($m_{\chi^\pm_1}>65\,{\rm GeV}$ if
$m_{\chi^\pm_1}-m_{\chi^0_1}>10\,{\rm GeV}$) and top-squark ($m_{\tilde t}$)
masses, limits from D0 on chargino masses, limits on Higgs boson masses, etc,
and run 365,000 SUSY models (for $1<\tan\beta\equiv{v_2\over v_1}<5$) and other
91,000 SUSY models (for $1<\tan\beta<1.5$) one finds~\cite{46}
\begin{equation}
\Delta R^{\rm susy}_b\le0.0017\ ,
\label{eq:15}
\end{equation}
not big enough to fill in the gap in (\ref{eq:13}). There are some recent
claims~\cite{47} that the upper bound (\ref{eq:15}) may be avoided in certain
very restrictive regions of parameter space that require a severe fine-tuning
of the parameters in the top-squark mass matrix. If then SUSY contributions
cannot make up the difference, what else is there to fix up $R_b$? As explained
in~\cite{10}, the existence of an extra light neutral gauge boson $Z'$, coupled
{\em only} to quarks, and {\em not} to leptons, thus {\em leptophobic}
\cite{48}, and with the right mixing with the regular $Z$, can do the job.
A leptophobic $Z'$ does not upset $\Gamma_{\rm lep}$, while by its appropriate
mixing ($\theta$) with the regular $Z$ it may be made phenomenologically
consistent with the SM, including $\Gamma_{\rm had}$. Actually, detailed fits
to the electroweak data~\cite{48,49,50} allow values for the parameter
$\delta\equiv\theta g_{Z'}/g_Z$ as large as $10^{-2}$. Even if an appropriate
$Z'$ may fit $R^{\rm exp}_b$, the standard question arises: ``who asked for
that?" Put it in a different way, if such a $Z'$ explanation is to be taken
seriously, one must provide a consistent theoretical framework where the new
gauge boson and its required properties arise naturally. Since a most important
theoretical issue, that of cancellation of gauge anomalies involving $Z'$, is
dealt with {\em automatically} in the string framework developed in the
previous section, there is where we have to look. New light neutral gauge
bosons ($Z'$) were early on considered to be the ``smoking guns" of string,
back when $E_6(\subset E_8)$ was the favorable, string-inspired gauge group
\cite{51}. It has been shown recently~\cite{48} that {\em dynamic leptophobia}
is possible via RGE U(1) mixing, and specifically the so-called $\eta$-model
\cite{52} in (string-inspired) $E_6$ stands out as the most reasonable model
that fits all the SM constraints and $R^{\rm exp}_b$. One may wonder if a more
natural way to achieve leptophobia exists, namely {\em symmetry-based
leptophobia}. Actually, we don't have to wonder very far since
SU(5)$\times$U(1) does the job~\cite{53}. Let me remind you that generically
in string theory, your ``chosen" (or ``preferred") gauge group is {\em
unavoidably} accompanied by extra U(1) factors. Sometimes it may be that the
extra U(1)'s are broken at the string scale, thus useless for providing light
$Z'$s. Nevertheless, it may happen that one extra U(1) survives unbroken down
to low energies, and if we are lucky, it may even fit the bill. Indeed, we
managed~\cite{53} to arrange our ``preferred" SU(5)$\times$U(1) model to be
accompanied by an extra U(1) surviving to low energies. What it rather stunning
is that leptophobia is very natural in SU(5)$\times$U(1). The reason is very
simple. A look at the way that quarks and leptons are distributed in the
{\bf10} and $\bar{\bf5}$ of SU(5)$\times$U(1), see (\ref{eq:11}), makes it
clear the fact that the {\bf10} contains {\em only} quarks (the $\nu^c$ is
superheavy), and the $\bar{\bf5}$ mixes quarks and leptons. Thus, it is very
easy to imagine a scheme where some $Z'$ couples only to {\bf10}'s and not
the $\bar{\bf5}$'s, thus ``leptophobic" because of symmetry reasons! Notice
that we cannot pull the same trick for ``canonical" SU(5) because, as is
apparent from (\ref{eq:12}), both {\bf10} and $\bar{\bf5}$ mix quarks and
leptons. String-based $Z'$ charge assignments lead to scenarios where $R_b$
is shifted significantly in the direction indicated experimentally, while
keeping $\Gamma_{\rm had}$ essentially unchanged and producing much smaller
shifts for $R_c$~\cite{53}. It is worth mentioning that, while the specific
$Z'$ couplings to quarks may change for different string realizations of
SU(5)$\times$U(1), there are certain phenomenological characteristics that
reflect the endemic SU(5)$\times$U(1) leptophobia, as quarks are largely split
from leptons in the SU(5) representations. Namely, maximal parity-violating
couplings to up-type quarks and parity-conserving couplings to down-type
quarks, that have the potential of yielding observable spin asymmetries in
polarized $pp$ scattering at RHIC~\cite{54}. The present experimental status
of $Z'$ gauge bosons was discussed here by Eppley and Wenzel~\cite{55}. For
a detailed phenomenological study of SU(5)$\times$U(1) $Z'$ bosons, including
production cross sections, additional contributions to the top-quark cross
section, and spin asymmetries at RHIC, see~\cite{53}. It is very important
to realize that even if $Z$-$Z'$ mixing {\em is not found} to be the resolution
of the $R_b$ puzzle, leptophobic $Z'$ gauge bosons may still be predicted by
string models (unmixed or negligibly mixed with the $Z'$) and their existence
should be probed experimentally in {\em all} possible ways. It is worth
emphasizing that even if the ``$R_{b,c}$-crisis" gets resolved purely
experimentally, something that looks, at least to me not inconceivable, the
would-be ``agreement" between $R^{\rm SM}_{b,c}$ and $R^{\rm exp}_{b,c}$ would
put severe constraints on possible extensions of the SM. In other words, the
``imagination stretch" now triggered by $R^{\rm exp}_{b,c}$ wouldn't have been
futile, because now ``all chips are down", e.g., (\ref{eq:15}), or possible
existence of leptophobic $Z'$ that can be probed experimentally in the near
future.
\subsection{The CDF $ee\gamma\gamma+E_T\hskip-13pt/\ $ ``event"}
As we heard from Carithers~\cite{56}, and further discussed in~\cite{10},
recent observations at the Tevatron, in the form of a puzzling
$ee\gamma\gamma+E_T\hskip-13pt/\ $ event~\cite{57}, appear to indicate that
experiment may have finally reach the sensitivity required to observe the
first {\em direct} manifestation of supersymmetry~\cite{58,59,44}. If this
event is indeed the result of an underlying supersymmetric production process,
as might be deduced from the observation of additional related events at the
Tevatron or LEP~2, then indeed we would have crossed a new threshold, literally
and methaphorically, in elementary particle physics. The particulars of the
event are listed in Table~\ref{Table1}.
\begin{table}[t]
\caption{The kinematical information of the observed CDF
$ee\gamma\gamma+E_T\hskip-13pt/\quad$ event. All momenta and energies in GeV.
Also important are $E_T\hskip-13pt/\quad=52.81\,{\rm GeV}$ at $\phi=2.91\,{\rm
rad}$.}
\label{Table1}
\begin{center}
\begin{tabular}{|crrrr|}\hline
Variable&$e_1\quad$&$e_2\quad$&$\gamma_1\quad$&$\gamma_2\quad$\\ \hline
$p_x$&$58.75$&$-33.41$&$-12.98$&$31.53$\\
$p_y$&$18.44$&$11.13$&$-29.68$&$-17.48$\\
$p_z$&$-167.24$&$21.00$&$-22.69$&$-34.77$\\
$E$&$178.21$&$41.00$&$39.55$&$50.09$\\
$E_T$&$61.58$&$35.21$&$32.39$&$36.05$\\ \hline
\end{tabular}
\end{center}
\end{table}
The direct evidence for supersymmetry contains the standard missing-energy
characteristic of supersymmetric production processes, but it also contains a
surprising {\em hard-photon} component (as far as minimalistic SUSY is
concerned), which eliminates all conceivable Standard Model backgrounds (e.g.,
if $WW\gamma\gamma$ is the origin of the ``event", less than $10^{-3}$ events
are expected with the current CDF data) and may prove extremely discriminating
among different models of low-energy supersymmetry. The present supersymmetric
explanations of the CDF event fall into two
phenomenological classes: either the lightest neutralino ($\chi^0_1$) is the
lightest supersymmetric particle, and the second-to-lightest neutralino decays
radiatively to it at the one-loop level ($\chi^0_2\to\chi^0_1\gamma$)
\cite{59}; or the gravitino ($\widetilde G$) is the lightest supersymmetric
particle, and the lightest neutralino decays radiatively to it at the tree
level ($\chi^0_1\to\widetilde G\gamma$)~\cite{58,59,44,60}. These two
explanations fall respectively into the ``minimalistic SUSY" framework
and the Very Light Gravitino (VLG) framework, discussed in Section 4.
The former ``neutralino-LSP" scenario requires~\cite{59} a configuration of
gaugino masses that precludes the usual gaugino mass unification relation of
unified models, although it can occur in some restricted region of the
``minimalistic SUSY" parameter space. The latter ``gravitino-LSP" scenario
requires only that the lightest neutralino has a photino component, as
is generically the case. The underlying process that leads to such final states
has been suggested to be that of selectron pair-production ($q\bar q\to
\widetilde e^+ \widetilde e^-$, $\widetilde e=\widetilde e_R,\widetilde e_L$),
with subsequent decay $\widetilde e\to e\chi^0_2$ or $\widetilde e\to
e\chi^0_1$ in the ``neutralino-LSP" and ``gravitino-LSP" scenarios
respectively. In the ``gravitino-LSP" scenario, the alternative possibility of
{\em chargino pair-production} ($q\bar q\to\chi^+_1\chi^-_1$,$\chi^\pm_1\to
e^\pm\nu_e\chi^0_1$) has also been suggested~\cite{60}.

Theoretically, the ``gravitino-LSP" explanation, belonging to the VLG framework
(see Section 4), is much more exciting and has generated model-building efforts
that try to embed such a scenario into a more fundamental theory at higher
mass scales. These more predictive theories include low-energy gauge-mediated
dynamical supersymmetry breaking~\cite{58,61} and no-scale supergravity
\cite{44}. In the former case (super)gravity seems to play a rather minuscule
role in the low-energy world, by essentially putting all the burden of SUSY
breaking into gauge (old or new) interactions. This sounds a little bizarre,
as one of the striking consequences of the VLG framework, as discussed in
Section 4, is the immense enhancement of the gravitational constant
($G_N\to G_N(\widetilde m/m_{3/2})^2$), in processes involving the would-be
goldstino, pushing it up to at least the Fermi constant
($G_F\sim10^{-5}m^2_{\rm proton}$)! On the other hand, in the latter case,
that of no-scale supergravity~\cite{44,60}, as discussed in the previous
section, (super)gravity plays a rather drastic role in the low-energy world,
and the aformentioned enhancement of the gravitational constant is used to
provide a {\em window of opportunity} to probe very high mass scales/very
short distances, including the exciting possibility of the unfolding of a
fifth-space dimension! As such, I will concentrate henceforth in the no-scale
supergravity interpretation~\cite{44,60}, since furthermore, the low-energy
gauge-mediated dynamical SUSY breaking has been covered in~\cite{10}.

The alert reader may have already noticed that the CDF ``event" is a striking
example of the VLG signature: $\gamma$'s plus missing energy
($E_T\hskip-13pt\quad\,$), as contained in the VLG one-parameter no-scale
supergravity~\cite{44}, that was developed in Section 4. Here is our strategy.
We delineated the regions in parameter space that are consistent with the
experimental kinematical information (see Table~\ref{Table1}), and then we
consider the rates for the various underlying processes that may occur within
such regions of parameter space. We also consider the constraints from LEP~1.5
and the prospects for SUSY particle detection at LEP161 and LEP190. The rather
restrictive nature of our one-parameter model make our experimental predictions
unambiguous and highly correlated. Here are our results.

\subsubsection{(A) Selectron interpretation}
The underlying process is $q\bar q\to\widetilde e^+_R\widetilde e^-_R$ or
$q\bar q\to\widetilde e^+_L\widetilde e^-_L$, with subsequent selectron decay
via $\widetilde e^\pm_{R,L}\to e^\pm\chi^0_1$ with 100\% B.R., followed by
neutralinos decaying via $\chi^0_1\to\gamma\widetilde G$, also with 100\% B.R.
The final state thus contains $e^+e^-\gamma\gamma\widetilde G\widetilde G$,
with the (essentially) massless gravitinos carrying away the missing energy.
We found that in our model~\cite{44,60} $\widetilde e_L$ production is
disfavored, while $\widetilde e_R$ is perfectly consistent with the kinematics
of the event, with
\begin{equation}
m_{\tilde e_R}\approx(85-135)\,{\rm GeV}\ ,
m_{\chi^0_1}\approx(50-100)\,{\rm GeV}\ {\rm implying}\
m_{\chi^\pm_1}\approx(90-190)\,{\rm GeV}
\label{eq:16}
\end{equation}
The specific microprocesses in the selectron interpretation are
\begin{eqnarray}
&&q\bar q\to \gamma,Z\to\widetilde\ell^+_R\widetilde\ell^-_R,\widetilde
\ell^+_L\widetilde\ell^-_L\to (\ell^+\chi^0_1)(\ell^-\chi^0_1)
\to \ell^+\ell^-\gamma\gamma+E_T\hskip-13pt/\quad\nonumber\\
&&q\bar q'\to W^\pm\to \widetilde \ell^\pm_L\widetilde\nu_\ell\to
(\ell^\pm\chi^0_1)(\nu_\ell\chi^0_1)
\to\ell^\pm\gamma\gamma+E_T\hskip-13pt/\quad,
\label{eq:17}\\
&&q\bar q\to Z\to\widetilde\nu_\ell\widetilde\nu_\ell\to
(\nu_\ell\chi^0_1)(\nu_\ell\chi^0_1)
\to\gamma\gamma+E_T\hskip-13pt/\quad,\nonumber
\end{eqnarray}
where $\ell=e,\mu,\tau$. Our calculations indicate~\cite{60} that at the
{\em one} dilepton-event level, the expected number of {\em single}-lepton
events (two) or {\em no}-lepton (diphoton) events (negligible) is still
consistent with observation (zero). However, possible observation of more
$ee\gamma\gamma+E_T\hskip-13pt/\quad$ events would need to be accompanied by
many more $e\gamma\gamma+E_T\hskip-13pt/\quad$ or
$\gamma\gamma+E_T\hskip-13pt/\quad$ events.

\subsubsection{(B) Chargino interpretation}
The underlying process is assumed to be $q\bar q\to\chi^+_1\chi^-_1$, with
subsequent chargino decay via $\chi^\pm_1\to e^\pm\nu_e\chi^0_1$ (with a
calculable B.R.), followed by the usual neutralino decay
$\chi^0_1\to\gamma\widetilde G$. The final state thus contains
$e^+e^-\gamma\gamma\nu_e\bar\nu_e\widetilde G\widetilde G$, with the
(essentially) massless gravitinos and the neutrinos carrying away the missing
energy. A similar analysis as above (A) shows that in this case one predicts
comparable rates (to the $ee\gamma\gamma+E_T\hskip-13pt/\quad$ event) for
$(\ell^\pm\ell^{'+}\ell^{'-},\ell^+\ell^-,\ell^+\ell^-jj)\,\gamma\gamma
+E_T\hskip-13pt/\quad$. In this interpretation it would be reasonable to
require $100\,{\rm GeV}<m_{\chi^\pm_1}<150\,{\rm GeV}$ (which implies
$m_{\tilde e_R}>85\,{\rm GeV}$ and $m_{\chi^0_1}>55\,{\rm GeV}$).

In the VLG--no-scale supergravity~\cite{44,60} interpretation (selectron or
chargino) of the CDF ``event", both selectrons and charginos (with masses
in the ${\cal O}(100\,{\rm GeV})$ region) seem to be kinematically inaccessible
at any LEP2 energy presently being considered (i.e.,
$\sqrt{s}=161,175,190$~GeV). One then has to focus on the $\chi^0_1\chi^0_1$
and perhaps also $\chi^0_1\chi^0_2$, as the only observable channels. In either
case an acoplanar photon pair plus missing energy
($\gamma\gamma+E_T\hskip-13pt/\quad$) final state, constitutes a rather clean
signal. The ``non-observation" of these events at LEP~1.5 does not exclude
any region of parameter space. It is worth mentioning that we have pointed out
\cite{60} that one of the acoplanar photon pairs observed by the OPAL
Collaboration~\cite{62} at LEP~1.5 may be attributable to supersymmetry in the
VLG--no-scale supergravity~\cite{44,60} model via $e^+e^-\to\chi^0_1\chi^0_1\to
\gamma\gamma+E_T\hskip-13pt/\quad$! Needless to say that we eagerly expect the
completion of the analyses of the current LEP run ($\sqrt{s}=161$~GeV).

\bigskip
It goes without saying that the consequences of confirming the supersymmetry
origin of the CDF ``event" will be rather dramatic. In at least one
interpretation it would not only provide evidence for a fundamental new
symmetry of Nature (supersymmetry), but it would also connect us directly,
through a fifth space dimension, to Planck-scale physics. On the other hand,
it may be that the CDF event is some sort of fluctuation/glitch/or whatever...
Hopefully, by the next workshop (XII) in this series, next year, we will know.

\section*{Acknowledgments}
It is a great pleasure to thank Luca Stanco and the other members of the
organizing committee for inviting me to give this summary talk, for their
warm hospitality, and for putting together a very exciting conference, that
among other things triggered my real interest in the CDF ``event". My thanks
are also due to Jorge Lopez for discussions and reading the manuscript.
This work has been supported in part by DOE grant DE-FG05-91-ER-40633.

\end{document}